\documentclass[12pt,a4paper]{article}
\usepackage{amsmath,amssymb,latexsym,epsfig,cite,color,graphics,graphicx,amsthm}
\usepackage{authblk}
\usepackage{braket}

\title{Polynomial Heisenberg algebras, \\[-4pt] multiphoton coherent states \\[-4pt] and geometric phases}

\author{Miguel Castillo-Celeita\footnote{mfcastillo@fis.cinvestav.mx}, Erik D\'iaz-Bautista\footnote{ediaz@fis.cinvestav.mx.\\
		Present address: Department of Theoretical Physics, Atomic Physics and Optics, University of Valladolid, 47011 Valladolid, Spain.} \\ and David J. Fern\'andez C.\footnote{david@fis.cinvestav.mx}}
\affil{\small Physics Department, Cinvestav, P.O. Box 14-740, 07000 Mexico City, Mexico}

\date{}

\begin{document}

\maketitle

\abstract{In this paper we will realize the polynomial Heisenberg algebras through the harmonic oscillator. We are 
going to construct then the Barut-Girardello coherent states, which coincide with the so-called multiphoton coherent states, and we will analyze the corresponding Heisenberg uncertainty relation and Wigner distribution function for some particular cases. We will show that these states are intrinsically quantum and cyclic, with a period being a 
fraction of the oscillator period. The associated geometric phases will be as well  evaluated.}

\section{Introduction}

Polynomial Heisenberg algebras (PHA) have become important deformations of the oscillator (Heisenberg-Weyl) 
algebra, that can be realized by finite-order differential ladder operators ${\cal L}^\pm$ which commute with the 
system Hamiltonian $H$ as the harmonic oscillator does, but between them giving place to an $m$-th degree polynomial 
$P_m(H)$ in $H$ \cite{m84,vs93,dek94,ad94,fhn94,as97,fh99,acin00,cfnn04}. The PHA are realized in a natural way by the SUSY partners of the 
harmonic and radial oscillators. Moreover, for degrees $2$ and $3$ they can be straightforwardly connected with 
the Painlev\'e IV and V equations, respectively. Thus, a simple method for generating solutions to these non-linear 
second-order ordinary differential equations has been recently designed \cite{cfnn04,bf14,bfn16}.

In order to determine the energy spectrum for systems ruled by PHA, it is of the highest importance to identify the 
extremal states, \textit{i.e.}, those states which are simultaneously eigenstates of the Hamiltonian and belong to 
the kernel of the annihilation operator ${\cal L}^-$. Once they have been found, departing from each one of them a 
ladder of energy eigenstates and eigenvalues (either finite or infinite) will be generated through the iterated action 
of the creation operator ${\cal L}^+$. The Hamiltonian spectrum then turns out to be the union of all these ladders of 
eigenvalues. However, for most of the systems studied until now the number of extremal states is less than the order 
of the differential annihilation operator. It would be important to know if there is a Hamiltonian for which the number 
of extremal states coincides with the order of the ladder operators which are involved. Moreover, we feel that it is 
time to identify the simplest system realizing in a non-trivial way the PHA. 

In this paper we will show that the system we are looking for is precisely the harmonic oscillator. Moreover, it will be 
seen that our algebraic method generalizes, thus it also includes, the standard approach. In order to implement our 
treatment, we will take the $k$-th power of the standard (first-order) annihilation and creation operators $a^\pm$ as the 
new ladder operators generating the PHA. This will cause that the Hilbert space ${\cal H}$ decomposes as a direct sum of 
$k$ orthogonal subspaces $\{{\cal H}_i, i=1,\dots,k\}$. In each ${\cal H}_i$ there will be just one extremal state, and 
by applying repeatedly the creation operator $(a^+)^k$ onto this state we will generate a ladder of new eigenstates, which 
will constitute the basis for this subspace.

Once the algebraic treatment has been completed, it will be natural to look for the Barut-Girardello coherent states of 
our system, as eigenstates of the annihilation operator with complex eigenvalue. We will study some important physical 
quantities for these states, such as the Heisenberg uncertainty relation and Wigner distribution function. We will see 
that these states coincide with the so-called multiphoton coherent states in the literature \cite{sp95,d02}, also called either crystallized Schr\"{o}dinger cat states \cite{clm95,cl12} or kaleidoscope coherent states \cite{pk18,kp18}. Moreover, it will be shown that they are 
intrinsically quantum cyclic states, with a period which is equal to the fraction $1/k$ of the oscillator period. As for 
any state evolving cyclically one can associate a {\it geometric phase}, which depends only of the geometry of the state 
space (the projective Hilbert space), we will calculate then such a phase for the cyclic motion performed by our CS.

This paper is organized as follows. In section 2 the Polynomial Heisenberg algebras will be briefly presented, while in 
section 3 we will realize them through the harmonic oscillator. We will describe as well the way of generating the energy 
spectrum and the eigenstates of the system by using PHA. In section 4 the multiphoton coherent states will be generated, 
as eigenstates of the $k$-th power of the annihilation operaror $a^-$, and some particular examples for these states will 
be analyzed, together with properties such as the Heisenberg uncertainty relation and Wigner distribution function. In the 
same section we will show that these states are cyclic and we will evaluate their associated geometric phase. Finally, in 
section 5 our conclusions will be presented.

\section{Polynomial Heisenberg algebras (PHA)}

Let us recall that the standard harmonic oscillator algebra, also known as Heisenberg-Weyl algebra, is generated by 
three operators $H$, $a^-$, $a^+$ which satisfy the following commutation relations:
\begin{subequations}
	\begin{eqnarray}
&& [H,a^\pm] = \pm a^\pm,\label{1} \\
&& [a^-,a^+] = 1, \label{2}
\end{eqnarray}
\end{subequations}
where the so-called number operator $N$ is linear in $H$, \textit{i.e.},
\begin{equation} \label{3}
N=a^+ \, a^- = H - \frac12.
\end{equation}

\begin{figure}
	\centering
	\includegraphics[scale=0.5]{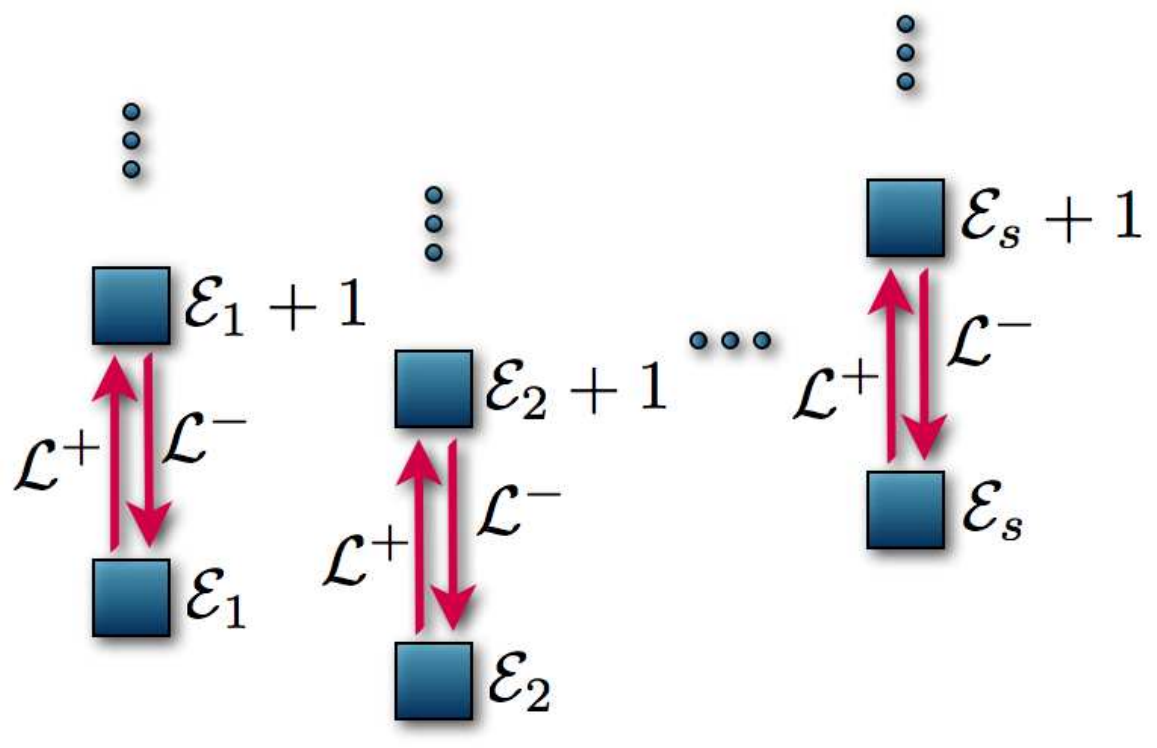}
	\caption{Energy spectrum for a system ruled by the PHA with $\omega=1$. The $s$ physical energy ladders start from the energy levels 
		${\cal E}_i, i = 1,\dots,s$, and each one of them characterizes the corresponding extremal state 
		$\psi_{{\cal E}_i}$.}\label{fig:ladders}
\end{figure}

On the other hand, the polynomial Heisenberg algebras of degree $m-1$ are deformations of the Heisenberg-Weyl 
algebra of kind:
\begin{subequations}
	\begin{eqnarray}
&& [H,{\cal L}^\pm] = \pm \omega \, {\cal L}^\pm \label{4} \\
&& [{\cal L}^-,{\cal L}^+] \equiv N_m(H + \omega ) - N_m(H)\equiv P_{m-1}(H), \label{5}
\end{eqnarray}
\end{subequations}
where the analogue of the number operator 
\begin{equation}\label{6}
N_m(H)\equiv {\cal L}^+ \, {\cal L}^-
\end{equation}
is a polynomial of degree $m$ in $H$.

It is possible to supply differential realizations of the PHA by assuming that $H$ is the following one-dimensional 
Schr\"odinger Hamiltonian
\begin{equation}\label{7}
H = -\frac12 \frac{d^2}{dx^2} + V(x),
\end{equation}
and ${\cal L}^\pm$ are differential ladder operators of order $m$ so that $N_m(H)$ is an $m$-th degree polynomial
in $H$, which can be factorized as
\begin{equation}\label{8}
N_m(H) = \prod_{i=1}^{m} \left(H - {\cal E}_i\right).
\end{equation}
Thus, $P_{m-1}(H)$ is a polynomial of degree $m-1$ in $H$.

The spectrum of $H$, ${\rm Sp}(H)$, can be generated by studying the Kernel $K_{{\cal L}^-}$ of ${\cal L}^-$ as 
follows
\begin{eqnarray}\label{9}
& {\cal L}^-\,\psi = 0 \quad \Rightarrow \quad
{\cal L}^+ {\cal L}^-\,\psi = \prod\limits_{i=1}^{m}\left(H - {\cal E}_i\right)\psi = 0 .
\end{eqnarray}
Since $K_{{\cal L}^-}$ in invariant under $H$,
\begin{equation}\label{10}
{\cal L}^- H \psi = (H+\omega){\cal L}^-\psi = 0 \quad \forall \ \ \psi \in K_{{\cal L}^-},
\end{equation}
then it is natural to take as a basis of $K_{{\cal L}^-}$ the states $\psi_{{\cal E}_i}$ such that
\begin{equation}\label{11}
H\psi_{{\cal E}_i} = {\cal E}_i \psi_{{\cal E}_i},
\end{equation}
which are called extremal states. From them we can generate, in principle, $m$ energy ladders with spacing $\Delta E = 
\omega$. However, if just $s$ extremal states satisfy the boundary conditions of the problem (we call them physical 
extremal states and order them as $\{\psi_{{\cal E}_i}, i = 1,\dots,s\}$), then from the iterated action of 
${\cal L}^+$ we will obtain $s$ physical energy ladders (see an illustration in Figure~\ref{fig:ladders}). We 
conclude that ${\rm Sp}(H)$ can have up to $m$ infinite ladders, with a spacing $\Delta E = \omega$ between steps.

\section{Polynomial Heisenberg algebras and the harmonic oscillator}
As mentioned previously, we can realize the PHA through systems ruled by one-dimensional Schr\"{o}dinger Hamiltonians. 
In particular, let us consider the harmonic oscillator, for which we will take $V(x)=\frac12x^2$ in Eq.~(\ref{7}). 
Moreover, let us construct the deformed ladder operators $a_g^-, a_g^+$ from the standard annihilation and creation 
operators $a^-, a^+$ as follows:
\begin{equation}\label{12}
a_g^- = (a^-)^k \qquad a_g^+ = (a^+)^k.
\end{equation}

The operator set $\{H, a_g^-, a_g^+\}$ generates a PHA of degree $k-1$, since it is fulfilled
\begin{subequations}
	\begin{eqnarray}
&& [H,a_g^\pm] =  \pm k a_g^\pm, \label{13}\\
&& [a_g^-,a_g^+] = N(H + k) - N(H), \label{14} \\
&& N(H) = a_g^+ \, a_g^- = \prod_{i=1}^{k} \left(H - i + \frac12\right). \label{15}
\end{eqnarray}
\end{subequations}
According to this formalism, we can identify now $k$ extremal state energies:
\begin{eqnarray}\label{16}
& {\cal E}_i = i - \frac12, \quad i=1,\dots,k,
\end{eqnarray}

\begin{figure}
	\centering
	\includegraphics[scale=1]{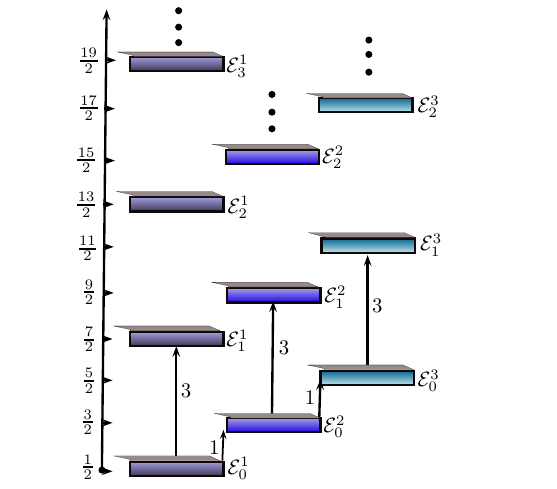}
\caption{The three independent energy ladders (with level spacing $\Delta E=3$) for the second-degree polynomial 
Heisenberg algebra of Eqs.~(\ref{13}-\ref{15}). Globally they reproduce the harmonic oscillator spectrum with the 
standard spacing $\Delta E=1$.}\label{fig:spectrum}
\end{figure}

\noindent so that the eigenvalues for the $i$-th ladder turn out to be:
\begin{eqnarray}\label{17}
& {\cal E}_{n}^i = {\cal E}_i + kn \quad n=0,1,\dots,
\end{eqnarray}
while the associated eigenstates are
\begin{eqnarray}\label{18}
& \vert \psi_{k,n}^i \rangle = \vert kn + i - 1 \rangle = 
\sqrt{\frac{(i-1)!}{(kn+i-1)!}}(a_g^+)^n \vert i - 1 \rangle.
\end{eqnarray}
According to the results of the previous section, the spectrum of the system becomes (see an example in 
Figure~\ref{fig:spectrum})
\begin{eqnarray}\label{19}
& {\rm Sp}(H) = \{{\cal E}_{0}^1, {\cal E}_{1}^1,\dots \} \cup 
\{{\cal E}_{0}^2, {\cal E}_{1}^2,\dots \}
\cup \dots \cup \{{\cal E}_{0}^k, {\cal E}_{1}^k,\dots \},
\end{eqnarray}
which is nothing but the harmonic oscillator spectrum, as it should be. In conclusion, we have addressed the 
harmonic oscillator based on an uncommon algebraic structure, the polynomial Heisenberg algebra, which 
nonetheless is associated in a natural way to this physical system. 

\section{Multiphoton coherent states}
After identifying the alternative algebraic structures underlying the harmonic oscillator, let us look for the 
corresponding coherent states. Although these states can be defined in several different ways, here we will build 
them as eigenstates of the deformed annihilation operator of Eq.~(\ref{12}), {\textit i.e.},
\begin{eqnarray}\label{20}
& a_g^- \vert \alpha \rangle_j = \alpha \vert \alpha \rangle_j,
\end{eqnarray}
where 
\begin{eqnarray}\label{21}
& \vert \alpha \rangle_j = \sum\limits_{n=0}^\infty C_n \vert kn + j \rangle, \quad j=0,\dots,k-1.
\end{eqnarray}
A standard calculation leads to the following normalized coherent states
\begin{eqnarray}\label{22}
\vert \alpha \rangle_j =
\frac{\vert\alpha\vert^{j/k}}{\left[H_{k,j}\left(\vert\alpha\vert^{2/k}\right)\right]^{1/2}}
\sum\limits_{n=0}^{\infty}\frac{\alpha^n}{\sqrt{(kn+j)!}}\vert kn + j \rangle,
\end{eqnarray}
where
\begin{equation}
H_{k,j}(x)=\sum_{m=0}^{\infty}\frac{x^{km+j}}{(km+j)!},
\end{equation}
is a generalized hyperbolic function of order $k$ and kind $j$ \cite{un82}.

Let us note that each coherent state $\vert \alpha \rangle_j$ is just a linear combination of the eigenstates of the 
harmonic oscillator Hamiltonian belonging to the $(j+1)$-th energy ladder, which generate the subspace ${\cal H}_{j+1}$. 
These coherent states are called multiphoton coherent states (MCS) in the literature, since they are superpositions 
of Fock states whose difference of energies is always an integer multiple of $k$ \cite{bg71,bjq90,b90}. This means that it is assumed that $k$ is the number of photons required to `jump' from the state $\vert kn + j \rangle$ 
to the next one $\vert k(n+1)+j\rangle$ in the $(j+1)$-th energy ladder. 

Let us analyze next some particular examples of these coherent states, for several values of $k$.

\subsection{Standard coherent states ($k=1$)}\label{4.1}
If we take $k=1$ and $j=0$ in Eq.~(\ref{22}), we recover the well known standard coherent states (SCS):
\begin{eqnarray}\label{23}
& \vert \alpha \rangle_0 =
\exp\left(-\frac{\vert \alpha\vert^2}2\right)
\sum\limits_{n=0}^{\infty}\frac{\alpha^n}{\sqrt{n!}}\vert n \rangle.
\end{eqnarray}
For these states it turns out that:
\begin{subequations}
	\begin{eqnarray}
\langle x\rangle_0 & = & \sqrt{2} \, {\rm Re}(\alpha),  \quad 
\langle p\rangle_0  =  \sqrt{2} \, {\rm Im}(\alpha), \label{24}  \\
\langle x^2\rangle_0 & = & \frac12 + \left[\sqrt{2} \, {\rm Re}(\alpha)\right]^2, \label{25} \\ 
\langle p^2\rangle_0 & = & \frac12 + \left[\sqrt{2} \, {\rm Im}(\alpha)\right]^2, \label{26} \\
(\Delta x)_0^2 & = & (\Delta p)_0^2 = (\Delta x)_0 \, (\Delta p)_0 = \frac12, \label{27}
\end{eqnarray}
\end{subequations}
where $x$ and $p$ are the position and momentum operator, respectively. This implies that the SCS are minimum 
uncertainty states.

\subsection{Multiphoton coherent states ($k>1$)}\label{4.2}
For $k\geq 2$ the mean value of the position and momentum operators, their squares and the Heisenberg uncertainty 
relation are given by:
\begin{subequations}
	\begin{eqnarray}
\langle x\rangle_j & = & \langle p\rangle_j = 0, \label{28} \\
\langle x^2\rangle_j & = & \vert a \vert \alpha \rangle_j \vert^2
+ \frac12 + {\rm Re}(\alpha)\delta_{k2}, \label{29} \\ 
\langle p^2\rangle_j & = & \vert a \vert \alpha \rangle_j \vert^2
+ \frac12 -{\rm Re}(\alpha)\delta_{k2}, \label{30} \\
(\Delta x)_j^2(\Delta p)_j^2 & = & \left(\vert a \vert \alpha \rangle_j \vert^2
+ \frac12\right)^2 - [{\rm Re}(\alpha)]^2 \delta_{k2}, \label{31} \\
\vert a\ket{{\alpha}}_{j}|^2&=&\frac{\vert\alpha\vert^{2/k}}{H_{k,j}\left(\vert\alpha\vert^{2/k}\right)}H_{k,k\delta_{0j}+j-1}\left(\vert\alpha\vert^{2/k}\right), \label{32}
\end{eqnarray}
\end{subequations}
where $\delta_{mn}$ is the Kronecker delta. In general, the MCS are not minimum uncertainty states, except for $j=0$ 
in the limit $\alpha \rightarrow 0$, since then $\vert \alpha \rangle_0 \rightarrow \vert 0\rangle$.

Another important quantity is the mean energy value for a system in a multiphoton coherent state, which turns out
to be
\begin{equation}\label{33}
\langle H \rangle_j =  \vert a \vert \alpha \rangle_j \vert^2
+ \frac12.
\end{equation}

On the other hand, in order to guarantee a {\it partial} completeness of the MCS in the subspace ${\cal H}_{j+1}$ to 
which $\vert \alpha \rangle_j$ belongs, {\it i.e.},
\begin{equation}\label{34}
\int \vert \alpha \rangle_j \, {}_j\langle \alpha \vert d \mu_j(\alpha) =  I_{j+1},
\end{equation}
with the measure $d \mu_j(\alpha)$ given by
\begin{equation}\label{35}
d \mu_j(\alpha) = \frac{1}{\pi}\frac{H_{k,j}\left(\vert\alpha\vert^{2/k}\right)}{\vert\alpha\vert^{2j/k+1}}
f_j(|\alpha|^2) d|\alpha| d \varphi,
\end{equation}
the function $f_j(x)$ must satisfy
\begin{equation}\label{36}
\int_0^\infty x^{n-1} f_j(x) dx = \Gamma(kn + j + 1).
\end{equation}
If Eqs.~(\ref{34}-\ref{36}) are fulfilled for any $j=0,\dots,k-1$, then it is valid a completeness relation in the 
full Hilbert space ${\cal H}$, since $I_1+ \dots+I_{k}= I$. This implies that any state can be decomposed 
in terms of the MCS.

Finally, the time evolution of a multiphoton coherent state becomes
\begin{equation}\label{37}
U(t)\vert \alpha \rangle_j =  \exp\left(-i(j+\frac12)t\right) \vert \alpha(t) \rangle_j , \quad \alpha(t) = \alpha \, \exp\left(-ikt\right).
\end{equation}
This equation reflects clearly the cyclic nature of the MCS, which have a period given by $\tau = 2\pi/k$ (the fraction 
$1/k$ of the harmonic oscillator period $T=2\pi$).

In order to clarify better these ideas, let us discuss explicitly two particular cases of multiphoton coherent states, 
which are known in the literature with given specific names.

\subsubsection{Biphoton coherent states with $k=2$}
For $k=2$ we get:
\begin{subequations}
	\begin{eqnarray}
&& \vert \alpha \rangle_0 =
\frac{1}{\left[H_{2,0}\left(\vert\alpha\vert\right)\right]^{1/2}}
\sum\limits_{n=0}^{\infty}\frac{\alpha^n}{\sqrt{(2n)!}}\vert 2n \rangle, \label{38} \\
&& \vert \alpha \rangle_1 =
\frac{\vert\alpha\vert^{1/2}}{\left[H_{2,1}\left(\vert\alpha\vert\right)\right]^{1/2}}
\sum\limits_{n=0}^{\infty}\frac{\alpha^n}{\sqrt{(2n+1)!}}\vert 2n + 1 \rangle. \label{39}
\end{eqnarray}
\end{subequations}
These states are called {\it even} and {\it odd} coherent states, respectively, since only even or odd energy eigenstates
of the oscillator are involved in the corresponding superposition \cite{mmkw96,bhdmmwrh96} (see also \cite{dmm74,xg89,dmn95,dkmm96,rr99,amaj16,cf16}).

For the states in Eqs.~(\ref{38}, \ref{39}), the expressions of Eqs.~(\ref{31}-\ref{33}) turn out to be:
\begin{subequations}
	\begin{eqnarray}
&&  \vert a \vert \alpha \rangle_0 \vert^2 = \vert \alpha \vert \tanh \vert \alpha \vert, \label{40} \\
&&  \vert a \vert \alpha \rangle_1 \vert^2 = \vert \alpha \vert \coth \vert \alpha \vert, \label{41} \\
&&  (\Delta x)_0^2(\Delta p)_0^2 = \left(\vert \alpha \vert \tanh \vert \alpha \vert
+ \frac12\right)^2 - [{\rm Re}(\alpha)]^2, \label{42} \\
&& (\Delta x)_1^2(\Delta p)_1^2 = \left(\vert \alpha \vert \coth \vert \alpha \vert
+ \frac12\right)^2 - [{\rm Re}(\alpha)]^2, \label{43} \\
&& \langle H \rangle_0 = \vert \alpha \vert \tanh \vert \alpha \vert + \frac12, \label{44} \\
&& \langle H \rangle_1 = \vert \alpha \vert \coth \vert \alpha \vert + \frac12. \label{45}
\end{eqnarray}
\end{subequations}

\begin{figure}
	\centering
\includegraphics[scale=0.7]{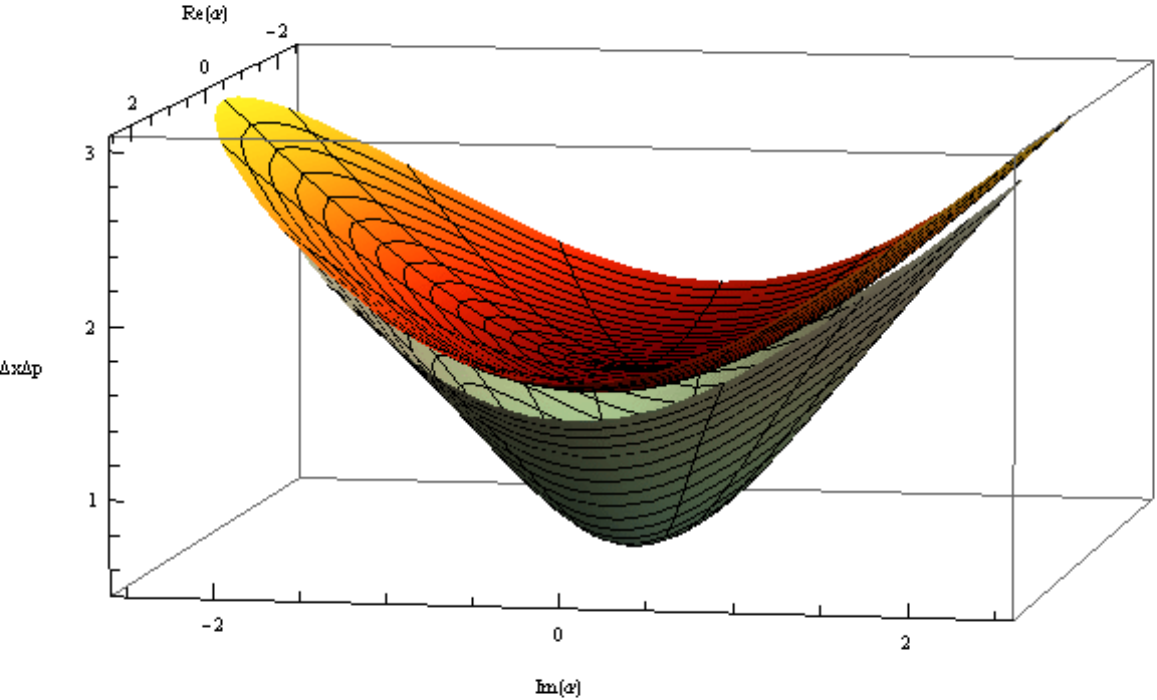}
\caption{Heisenberg uncertainty relation $(\Delta x)_j(\Delta p)_j$, $j=0,1$ as a function of $\alpha$. In the 
limit $\alpha\rightarrow0$ the uncertainty relation for the {\it even} coherent states reaches its lowest possible 
value ($\frac12$) while for the {\it odd} states it just acquires the local minimum, equal to $\frac32$.}
\label{fig:unceraintyk2}
\end{figure}

It is straightforward to check that the {\it even} states are minimum uncertainty states for 
$\alpha=0$, while in general the {\it odd} states do not have this property (see Figure \ref{fig:unceraintyk2}).

\subsubsection{Triphoton coherent states with $k=3$}
By making now $k=3$ we have \cite{sww91,cf17}:
\begin{subequations}
	\begin{eqnarray}
&& \vert \alpha \rangle_0 =
\frac{1}{\left[H_{3,0}\left(\vert\alpha\vert^{2/3}\right)\right]^{1/2}}
\sum\limits_{n=0}^{\infty}\frac{\alpha^n}{\sqrt{(3n)!}}\vert 3n \rangle, \label{46} \\
&& \vert \alpha \rangle_1 =
\frac{\vert\alpha\vert^{1/3}}{\left[H_{3,1}\left(\vert\alpha\vert^{2/3}\right)\right]^{1/2}}
\sum\limits_{n=0}^{\infty}\frac{\alpha^n}{\sqrt{(3n+1)!}}\vert 3n+1 \rangle, \label{47} \\
&& \vert \alpha \rangle_2 =
\frac{\vert\alpha\vert^{2/3}}{\left[H_{3,2}\left(\vert\alpha\vert^{2/3}\right)\right]^{1/2}}
\sum\limits_{n=0}^{\infty}\frac{\alpha^n}{\sqrt{(3n+2)!}}\vert 3n+2 \rangle. \label{48}
\end{eqnarray}
\end{subequations}

\begin{figure}
	\centering
	\includegraphics[scale=0.7]{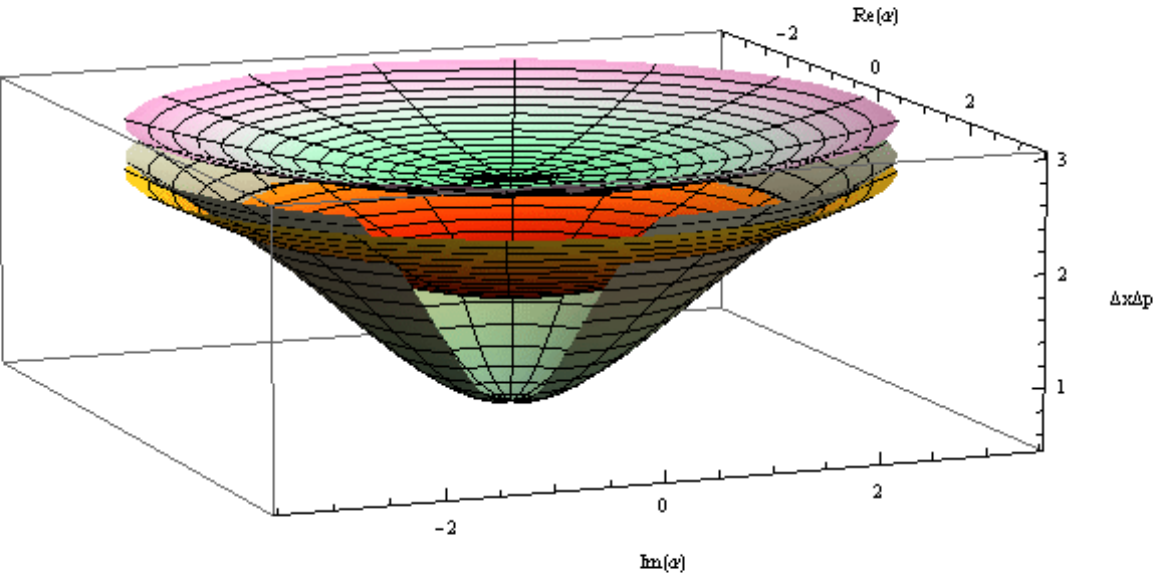}
	\caption{Heisenberg uncertainty relation $(\Delta x)_j(\Delta p)_j$, $j=0,1,2$ as function of $\alpha$. When $\alpha\rightarrow0$ the triphoton coherent state with $j=0$ reaches the lowest possible value ($\frac12$) while the ones with $j=1$ and $j=2$ just acquire the local minimum, equal to $\frac32$ and $\frac52$, respectively.}\label{fig:unceraintyk3}
\end{figure}

\noindent We call these states the {\it good}, the {\it bad} and the {\it ugly} coherent states, respectively \cite{cf17}; they have been also called {\it trinity states} in the literature \cite{pk18,kp18}.

Similarly to the previous case, the Heisenberg uncertainty relation and mean energy value for these states are obtained 
by taking $k=3$ in Eqs.~(\ref{31}-\ref{33}), which leads to:
	\begin{equation}\label{49}
(\Delta x)_j(\Delta p)_j= \langle H\rangle_j=\vert a\vert \alpha\rangle_j\vert^2+\frac12, \quad j=0,1,2,
\end{equation}
where
\begin{subequations}
\begin{align}
\nonumber&  \vert a \vert \alpha \rangle_0 \vert^2 = \frac{1}{H_{3,0}\left(\vert\alpha\vert^{2/3}\right)}
\sum\limits_{n=0}^{\infty}\frac{|\alpha|^{2n+2}}{(3n+2)!}  \\
& \quad =\vert\alpha\vert^{2/3}\left[e^{\frac{3\vert\alpha\vert^{2/3}}{2}}+2 \cos \left(\frac{\sqrt{3} \vert\alpha\vert^{2/3}}{2}\right)\right]^{-1}\left[e^{\frac{3\vert\alpha\vert^{2/3}}{2}}-2 \sin \left(\frac{\pi}{6}+\frac{\sqrt{3} \vert\alpha\vert^{2/3}}{2} \right)\right], \label{50} \\
\nonumber&  \vert a \vert \alpha \rangle_1 \vert^2 = \frac{\vert\alpha\vert^{2/3}}{H_{3,1}\left(\vert\alpha\vert^{2/3}\right)}
\sum\limits_{n=0}^{\infty}\frac{|\alpha|^{2n}}{(3n)!} \\
& \quad = \vert\alpha\vert^{2/3}\left[e^{\frac{3\vert\alpha\vert^{2/3}}{2}}-2 \sin \left(\frac{\pi}{6} -\frac{\sqrt{3} \vert\alpha\vert^{2/3}}{2}\right)\right]^{-1}\left[e^{\frac{3\vert\alpha\vert^{2/3}}{2}}+2 \cos \left(\frac{\sqrt{3} \vert\alpha\vert^{2/3}}{2}\right)\right] \label{51} \\
\nonumber&  \vert a \vert \alpha \rangle_2 \vert^2 = \frac{\vert\alpha\vert^{4/3}}{H_{3,2}\left(\vert\alpha\vert^{2/3}\right)}
\sum\limits_{n=0}^{\infty}\frac{|\alpha|^{2n}}{(3n+1)!} \\
& \quad =\vert\alpha\vert^{2/3}\left[e^{\frac{3\vert\alpha\vert^{2/3}}{2}}-2 \sin \left(\frac{\pi}{6}+\frac{\sqrt{3}}{2} \vert\alpha\vert^{2/3} \right)\right]^{-1}\left[e^{\frac{3\vert\alpha\vert^{2/3}}{2}}-2 \sin \left(\frac{\pi}{6}-\frac{\sqrt{3}}{2} \vert\alpha\vert^{2/3}\right)\right]. \label{52}
\end{align}
\end{subequations}

We can see that in the vicinity of $\alpha=0$ the uncertainty relation for the triphoton coherent state with $j=0$ achieves the lowest possible value, while for the other ones (with $j=1$ and $j=2$) it just acquires a local minimum (see  Figure~\ref{fig:unceraintyk3}).

\subsection{Superposition of standard coherent states}\label{4.3}

Let us consider now the following unnormalized CS:
\begin{subequations}
	\begin{eqnarray}
&& \vert z\rangle = \sum_{n=0}^\infty \frac{z^n}{\sqrt{n!}} \vert n\rangle, \label{53} \\
&& \vert z\rangle_j = \sum_{n=0}^\infty \frac{z^{kn+j}}{\sqrt{(kn+j)!}} \vert kn+j\rangle, 
\quad \alpha = z^k. \label{54}
\end{eqnarray}
\end{subequations}

The states of Eq.~(\ref{53}) are the (unnormalized) standard coherent states of section~4.1 while those of Eq.~(\ref{54}) 
are the multiphoton coherent states of section~4.2, in which we have taken $\alpha=z^k$. It is clear that the SCS are just a particular case of the MCS for $k=1$, $j=0$ (see subsection \ref{4.1}).

According to the completeness relationship in Eq.~(\ref{34}), and the related discussion, any state in ${\cal H}$ can be 
decomposed in terms of the states $\vert z\rangle_j, \ j=0,\dots,k-1$.

\begin{figure}
	\centering
	\includegraphics[scale=1]{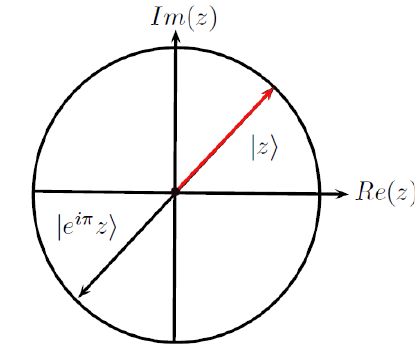}
	\caption{Diagram representing the two standard coherent states $\vert z\rangle$ and $\vert e^{i\pi}z\rangle$ in the complex plane which give place to the {\it even} and {\it odd} coherent states $\vert z\rangle_0$ and 
	$\vert z\rangle_1$.}\label{fig:circlek2}
\end{figure}

In particular, for $k=2$ let us consider the standard coherent states $\vert z\rangle$ and $\vert e^{i\pi}z\rangle$, whose 
eigenvalues have a phase difference of $\pi$ \cite{d02,di06,mtk90}. It is straightforward to verify that:
\begin{subequations}
	\begin{eqnarray}
&& \vert z\rangle = \vert z\rangle_0 + \vert z\rangle_1, \label{55} \\
&& \vert e^{i\pi}z\rangle = \vert z\rangle_0 - \vert z\rangle_1. \label{56}
\end{eqnarray}
\end{subequations}

From these equations we can solve now the {\it even} and {\it odd} coherent states in terms of the two standard ones, 
$\vert z\rangle$ and $\vert e^{i\pi}z\rangle$. After normalization we get that:
\begin{subequations}
	\begin{eqnarray}
&& \vert z\rangle_0 = \frac{e^{-|z|^2/2}}{\sqrt{2(1+e^{-2|z|^2})}}[\vert z\rangle + \vert e^{i\pi}z\rangle], \label{57} \\
&& \vert z\rangle_1 = \frac{e^{-|z|^2/2}}{\sqrt{2(1-e^{-2|z|^2})}}[\vert z\rangle - \vert e^{i\pi}z\rangle], \label{58}
\end{eqnarray}
\end{subequations}
which means that the states $\vert z\rangle_0$ and $\vert z\rangle_1$ are linear combinations of two standard coherent  states with opposite positions in the complex plane (see Figure \ref{fig:circlek2}). Note that expressions in equations (\ref{55}-\ref{58}) are called quantum Fourier transforms \cite{pk18}.

\begin{figure}
	\centering
\includegraphics[scale=0.7]{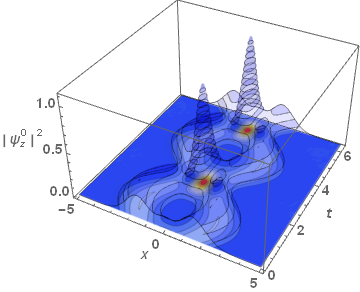}
\caption{Probability density $|\psi_z^0(x,t)|^2$ for an {\it even} coherent state.}\label{fig:psik2j0}
\end{figure}
\begin{figure}
	\centering
	\includegraphics[scale=0.7]{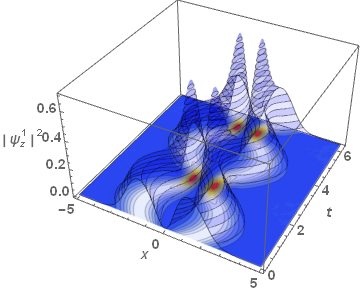}
	\caption{Probability density $|\psi_z^1(x,t)|^2$ for an {\it odd} coherent state.}\label{fig:psik2j1}
\end{figure}

We can use now the wavefunction of a normalized standard coherent state in order to analyze the time evolution of the {\it even} and {\it odd} coherent states. First of all we have
\begin{equation}\label{59}
\psi_z(x)=\langle x\vert z\rangle=\left(\frac{1}{\pi}\right)^{1/4}e^{-\frac12(x-\langle x\rangle)^2+ix\langle p\rangle},
\end{equation}
with $\langle x\rangle$ and $\langle p\rangle$ as given in Eq.~(\ref{24}). Thus, the wavefuntions associated to the states in 
Eqs.~(\ref{57}-\ref{58}) become \cite{r13}:
\begin{subequations}
	\begin{eqnarray}
&& \psi^0_z(x)=\langle x\vert z\rangle_0=N_0\left[e^{-\frac12(x-\langle x\rangle)^2+ix\langle p\rangle}+e^{-\frac12(x+\langle x\rangle)^2-ix\langle p\rangle}\right], \label{60} \\
&& \psi^1_z(x)=\langle x\vert z\rangle_1=N_1\left[e^{-\frac12(x-\langle x\rangle)^2+ix\langle p\rangle}-e^{-\frac12(x+\langle x\rangle)^2-ix\langle p\rangle}\right], \label{61}
\end{eqnarray}
\end{subequations}
where
\begin{equation}\label{62}
N_0=\left(\frac{1}{\pi}\right)^{1/4}\frac{1}{\sqrt{2(1+e^{-\langle x\rangle^2-\langle p\rangle^2})}}, \quad N_1=\left(\frac{1}{\pi}\right)^{1/4}\frac{1}{\sqrt{2(1-e^{-\langle x\rangle^2-\langle p\rangle^2})}}.
\end{equation}

The time-dependent wavefunctions $\psi^j_z(x,t)=\langle x\vert U(t)\vert z\rangle_j$, $j=0,1$ appear by replacing in 
Eqs.~(\ref{60}-\ref{61}) $\langle x\rangle\rightarrow\langle x\rangle \cos t+\langle p\rangle \sin t$ and 
$\langle p\rangle\rightarrow\langle p\rangle \cos t-\langle x\rangle \sin t$. In Figures \ref{fig:psik2j0} and \ref{fig:psik2j1} 
we illustrate the corresponding probability densities for $j=0$ and $j=1$, respectively, for the eigenvalue $z=1+i$.

It is important to stress that the coherent states $U(t)\vert z\rangle_0$ and $U(t)\vert z\rangle_1$ are cyclic, with a period 
$\tau=\pi$ which is one half of the oscillator period $T\equiv2\pi$. This fact implies that both, the {\it even} and 
{\it odd} coherent states are intrinsically quantum, since they recover their initial form before the oscillator period has 
elapsed \cite{cf16}.

On the other hand, for $k=3$ we consider the three standard coherent states whose eigenvalues posses a phase difference of 
$2\pi/3$, i.e., $\vert z\rangle$, $\vert e^{i2\pi/3}z\rangle$ and $\vert e^{i4\pi/3}z\rangle$. First we express these states in 
terms of the triphoton coherent states as follows: 
\begin{subequations}
	\begin{eqnarray}
&& \vert z\rangle = \vert z\rangle_0 + \vert z\rangle_1 + \vert z\rangle_2, \label{63} \\
&& \vert e^{i2\pi/3}z\rangle = \vert z\rangle_0 + e^{i2\pi/3}\vert z\rangle_1 + e^{i4\pi/3} \vert z\rangle_2, \label{64} \\
&& \vert e^{i4\pi/3}z\rangle = \vert z\rangle_0 + e^{i4\pi/3}\vert z\rangle_1 + e^{i8\pi/3} \vert z\rangle_2. \label{65}
\end{eqnarray}
\end{subequations}

As before, from these expressions we can solve the eigenstates $\vert z\rangle_0, \ \vert z\rangle_1, \ \vert z\rangle_2$ of 
the annihilation operator $a^-_g=(a^-)^3$ in terms of the standard coherent states $\vert z\rangle$, $\vert e^{i2\pi/3}z\rangle$ 
and $\vert e^{i4\pi/3}z\rangle$:
\begin{subequations}
	\begin{eqnarray}
&& \vert z\rangle_0 = N_0[\vert z\rangle + \vert e^{i2\pi/3}z\rangle + \vert e^{i4\pi/3}z\rangle], \label{66} \\
&& \vert z\rangle_1 = N_1[\vert z\rangle - e^{i\pi/3}\vert e^{i2\pi/3}z\rangle + e^{i2\pi/3}\vert e^{i4\pi/3}z\rangle], \label{67} \\
&& \vert z\rangle_2 = N_2[\vert z\rangle + e^{i2\pi/3}\vert e^{i2\pi/3}z\rangle + e^{i4\pi/3}\vert e^{i4\pi/3}z\rangle], \label{68}
\end{eqnarray}
\end{subequations}
where $N_j$, $j=0,1,2$ are normalization constants. We note once again that expression in equations (\ref{63}-\ref{68}) are called quantum Fourier transforms.

\begin{figure}
	\centering
	\includegraphics[scale=1]{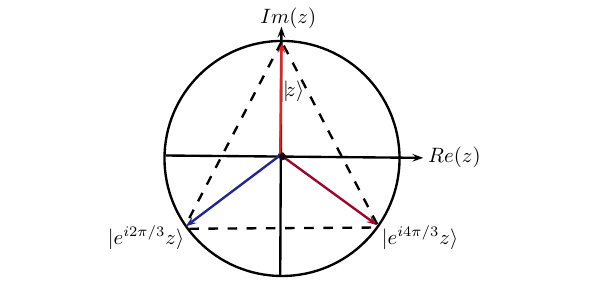}
	\caption{Diagram showing the position in the complex plane of the standard coherent states $\vert z\rangle$, $\vert e^{i2\pi/3}z\rangle$ 
	and $\vert e^{i4\pi/3}z\rangle$, whose superpositions give place to the triphoton coherent states $\vert z\rangle_0$, $\vert z\rangle_1$ and $\vert z\rangle_2$.}\label{fig:circlek3}
\end{figure}

Note that  the states $\vert z\rangle_0$, $\vert z\rangle_1$ and $\vert z\rangle_2$ are now linear combinations of three 
standard coherent states which are placed in the vertices of an equilateral triangle in the complex plane 
(see Figure~\ref{fig:circlek3}).

Once again, using the wavefunction of Eq.~(\ref{59}) we can get the wavefunctions associated to the states in 
Eqs.~(\ref{66}-\ref{68}):
\begin{subequations}
	\begin{eqnarray}
\psi^0_z(x)&=&N_0 \left[e^{-\frac12(x-\langle x\rangle)^2+ix\langle p\rangle}+e^{-\frac12(x-\langle x\rangle_1)^2+ix\langle p\rangle_1}+e^{-\frac12(x-\langle x\rangle_2)^2+ix\langle p\rangle_2}\right], \label{69}\\
\nonumber\psi^1_z(x)&=&N_1 \left[e^{-\frac12(x-\langle x\rangle)^2+ix\langle p\rangle}-e^{i\pi/3}e^{-\frac12(x-\langle x\rangle_1)^2+ix\langle p\rangle_1}\right.\\
&&\left.+e^{i2\pi/3}e^{-\frac12(x-\langle x\rangle_2)^2+ix\langle p\rangle_2}\right], \label{70}\\
\nonumber\psi^2_z(x)&=&N_2 \left[e^{-\frac12(x-\langle x\rangle)^2+ix\langle p\rangle}+e^{2i\pi/3}e^{-\frac12(x-\langle x\rangle_1)^2+ix\langle p\rangle_1}\right.\\
&&\left.+e^{i4\pi/3}e^{-\frac12(x-\langle x\rangle_2)^2+ix\langle p\rangle_2}\right], \label{71}
\end{eqnarray}
\end{subequations}
where $\langle x\rangle$ and $\langle p\rangle$ are given in Eq.~(\ref{24}) while
\begin{subequations}
	\begin{eqnarray}
&& \langle x\rangle_1=\sqrt{2}{\rm Re}(ze^{i2\pi/3}), \quad \langle p\rangle_1=\sqrt{2}{\rm Im}(ze^{i2\pi/3}), \label{72}  \\
&& \langle x\rangle_2=\sqrt{2}{\rm Re}(ze^{i4\pi/3}), \quad \langle p\rangle_2=\sqrt{2}{\rm Im}(ze^{i4\pi/3}), \label{73}
\end{eqnarray}
\end{subequations}
and
\begin{subequations}
	\begin{eqnarray}
\nonumber N_0&=&\left(\frac{1}{\pi}\right)^{1/4}\frac13\left|e^{\vert z\vert^2}+e^{\vert z\vert^2e^{i2\pi/3}}+e^{\vert z\vert^2e^{i4\pi/3}}\right|^{-1/2} \\
&=&\left(\frac{1}{\pi}\right)^{1/4}\frac13\left[\frac{1}{3} \left(e^{\vert z\vert^2}+2 e^{-\frac{\vert z\vert^2}{2}} \cos \left(\frac{\sqrt{3} \vert z\vert^2}{2}\right)\right)\right]^{-1/2}, \label{74} \\
\nonumber N_1&=&\left(\frac{1}{\pi}\right)^{1/4}\frac13\left|e^{\vert z\vert^2}+e^{i\pi/3}e^{\vert z\vert^2e^{i2\pi/3}}+e^{i2\pi/3}e^{\vert z\vert^2e^{i4\pi/3}}\right|^{-1/2} \\
&=&\left(\frac{1}{\pi}\right)^{1/4}\frac13\left[\frac{1}{3}\left(e^{\vert z\vert^2}-2 e^{-\frac{\vert z\vert^2}{2}} \sin \left(\frac{\pi}{6} -\frac{\sqrt{3} \vert z\vert^2}{2}\right)\right)\right]^{-1/2}, \label{75} \\
\nonumber N_2&=&\left(\frac{1}{\pi}\right)^{1/4}\frac13\left|e^{\vert z\vert^2}+e^{i2\pi/3}e^{\vert z\vert^2e^{i2\pi/3}}+e^{i4\pi/3}e^{\vert z\vert^2e^{i4\pi/3}}\right|^{-1/2} \\
&=& \left(\frac{1}{\pi}\right)^{1/4}\frac13\left[\frac{1}{3}\left(e^{\vert z\vert^2}-2 e^{-\frac{\vert z\vert^2}{2}} \sin \left(\frac{\pi}{6}+\frac{\sqrt{3} \vert z\vert^2}{2} \right)\right)\right]^{-1/2}. \label{76}
\end{eqnarray}
\end{subequations}

\begin{figure}
	\centering
\includegraphics[scale=0.7]{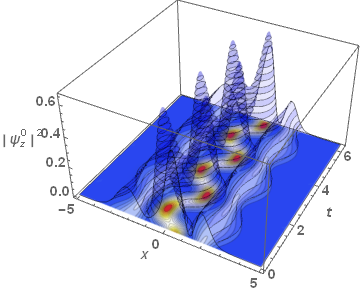}
\caption{Probability density $|\psi_z^0(x,t)|^2$ for a triphoton coherent state with $j=0$.}\label{fig:psik3j0}
\end{figure}

\begin{figure}
	\centering
\includegraphics[scale=0.7]{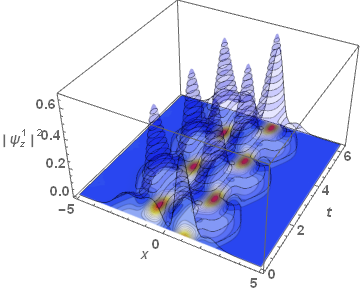}
\caption{Probability density $|\psi_z^1(x,t)|^2$ for a triphoton coherent state with $j=1$.}\label{fig:psik3j1}
\end{figure}

\begin{figure}
	\centering
\includegraphics[scale=0.7]{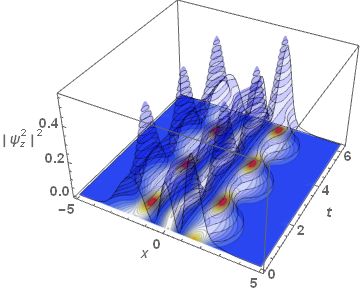}
\caption{Probability density $|\psi_z^2(x,t)|^2$ for an triphoton coherent state with $j=2$.}\label{fig:psik3j2}
\end{figure}

By applying now the evolution operator $U(t)$ on the states of Eqs.~(\ref{66}-\ref{68}) we can obtain the corresponding 
time-dependent wavefunctions $\psi^j_z(x,t)=\langle x\vert U(t)\vert z\rangle_j$, $j=0,1,2$. Examples of the corresponding 
probability densities for the eigenvalue $z=1+i$ are shown in Figure \ref{fig:psik3j0} for $j=0$, in Figure \ref{fig:psik3j1} for $j=1$ and in 
Figure~\ref{fig:psik3j2} for $j=2$.

As in the previous case, the evolving coherent states $U(t)\vert z\rangle_0$, $U(t)\vert z\rangle_1$ and $U(t)\vert z\rangle_2$ 
are also cyclic, but with a period $\tau=T/3=2\pi/3$. This means that these states recover their initial form before the 
oscillator period has elapsed, {\it i.e.}, the triphoton coherent states exhibit a marked quantum behavior.

\begin{figure}[htbp]
	\centering
	{\includegraphics[width=150mm]{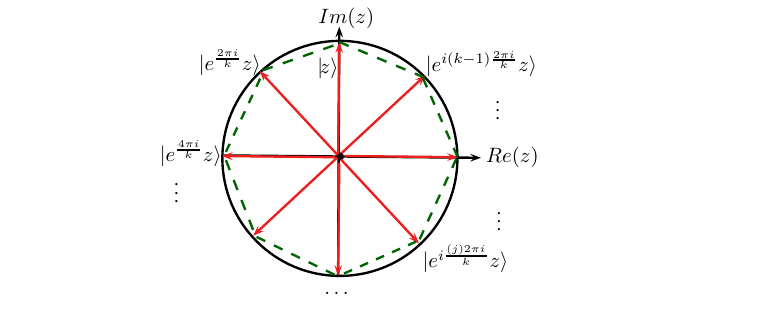}}
	\caption{Graphic representation of the $k$ SCS in the complex plane leading to the MCS $\vert z\rangle_j$, $j=0,\dots,k-1$. The associated symmetry in the one of a regular polygon (dashed line).} \label{KOCS}
\end{figure}

For arbitrary $k$, the process is analogous to the second and third order cases. Firstly, the SCS which are rotated subsequently by a phase $\mu=\exp\left(\frac{2\pi i}{k}\right)$ are expanded in terms of the MCS as $\ket{\mu^jz}=\mathbf{M}_{j,l}\ket{z}_l$, with the coefficients of the expansion constituting the cyclic group, as it is explored in~\cite{clm95,cl12}:

\begin{equation}
	\begin{bmatrix}
		\ket{z}\\
		\ket{\mu z}\\
		\ket{\mu^2 z}\\
		\vdots\\
		\ket{\mu^{(k-1)}z}
	\end{bmatrix} 
	=
	\begin{bmatrix}
		1 & 1 & 1 &\ldots& 1 \\
		1 & \mu & \mu^2 &\ldots& \mu^{(k-1)} \\
		1 & \mu^2 & \mu^{2(2)} &\ldots& \mu^{2(k-1)} \\
		\vdots&\vdots&\vdots&\vdots&\vdots\\
		1 & \mu^{(k-1)} & \mu^{2(k-1)} &\ldots& \mu^{(k-1)^2} \\
	\end{bmatrix}
	\begin{bmatrix}
		\ket{z}_0\\
		\ket{z}_1\\
		\ket{z}_2\\
		\vdots\\
		\ket{z}_{k-1}
	\end{bmatrix}.
\end{equation}
The SCS which are involved can be represented in the complex plane as in Fig~(\ref{KOCS}).

In the second place, we express the MCS in terms of the SCS $\ket{\mu^jz}=\mathbf{M}^{-1}_{j,l}\ket{z}_l$, where the inverse $\mathbf{M}_{j,l}^{-1}$ is calculated through standard methods

\begin{equation}
\begin{bmatrix}
\ket{z}_0\\
\ket{z}_1\\
\ket{z}_2\\
\vdots\\
\ket{z}_{k-1}
\end{bmatrix}
=
\frac{1}{k}\begin{bmatrix}
1 & 1 & 1 &\ldots& 1 \\
1 & \mu^{-1} & \mu^{-2} &\ldots& \mu^{-(k-1)} \\
1 & \mu^{-2} & \mu^{-2(2)} &\ldots& \mu^{-2(k-1)} \\
\vdots&\vdots&\vdots&\vdots&\vdots\\
1 & \mu^{-(k-1)} & \mu^{-2(k-1)} &\ldots& \mu^{-(k-1)^2} \\
\end{bmatrix}
\begin{bmatrix}
\ket{z}\\
\ket{\mu z}\\
\ket{\mu^2 z}\\
\vdots\\
\ket{\mu^{(k-1)}z}
\end{bmatrix}.
\end{equation}
In this way we obtained the expressions given in Eqs.~(\ref{57},\ref{58}) and Eqs.~(\ref{66}-\ref{68}). These decompositions guarantee to work with the SCS wave functions, which are more familiar than the ones for the multiphoton coherent states.

Also, the above expression allows to find the wavefunction of the multiphoton coherent states for arbitrary $k$ as:
	\begin{equation}\label{76.1}
	\psi_z^j(x)=N_j\sum_{s=0}^{k-1}\mu^{-sj}\psi_{z\mu^{s}}(x), \quad j=0,1,\dots,k-1,
	\end{equation}
	where the factor $1/k$ has been absorbed in the normalization constants $N_j$.

Let us note that the multiphoton coherent states studied here supply an interesting representation of a quantum symmetry related with the quantum $q$-oscillator \cite{pk18}.

\subsection{Wigner distribution function}
As is well known \cite{w32}, the Wigner distribution function for a system in a state $\vert\psi\rangle$ is given by
\begin{align}
\nonumber W(q,p) \equiv & \frac{1}{2\pi}\int_{-\infty}^{\infty} \langle q - \frac{y}{2}\vert\psi\rangle
\langle\psi \vert q + \frac{y}2 \rangle \exp(ipy) dy \\
= & \frac{1}{\pi}\int_{-\infty}^{\infty} \psi^*(q+y) \psi(q-y) \exp(2ipy) dy, \label{77}
\end{align}
which is a quasi-probability density since
\begin{subequations}
	\begin{eqnarray}
\int_{-\infty}^{\infty}W(q,p) dp = & \vert\psi(q)\vert^2, \label{78} \\
\int_{-\infty}^{\infty}W(q,p) dq = & \vert\varphi(p)\vert^2, \label{79}
\end{eqnarray}
\end{subequations}
but it is not a positive definite function. In fact, the Wigner function can take negative values, a property that can be used as a sign of the quantumness of a state \cite{kz04}. This fact will be used now to analyze the multiphoton coherent states.

\begin{figure}
	\centering
\includegraphics[scale=0.7]{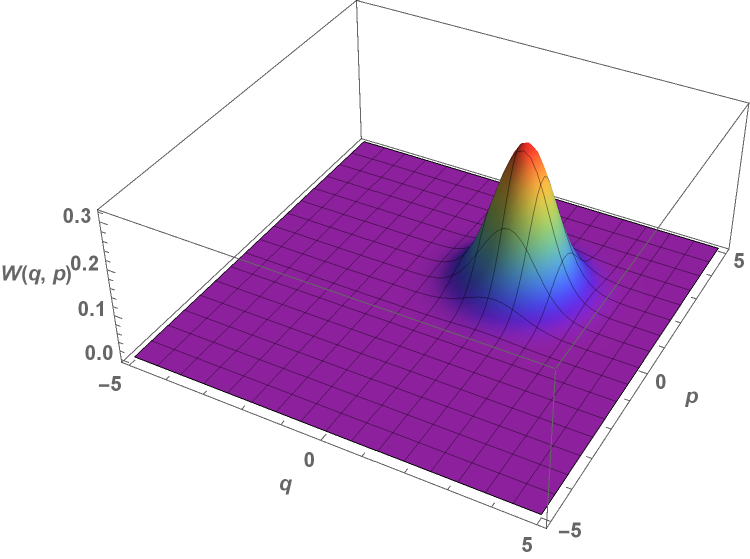}
\caption{Wigner distribution function $W(q,p)$ for a standard coherent state which has a Gaussian form, as for the harmonic oscillator ground state, but now centered in the point $(q_0,p_0)=(\sqrt{2}{\rm Re}(z),\sqrt{2}{\rm Im}(z))$ in 
phase space.}\label{fig:W10}
\end{figure}

The Wigner function associated to the SCS is obtained by substituing the wavefunction of Eq.~(\ref{59}) in Eq.~(\ref{77}),
leading to:
\begin{equation}\label{80}
W_z(q,p)=\frac1\pi e^{-(q-\langle q\rangle)^2}e^{-(p-\langle p\rangle)^2},
\end{equation}
where $\langle q\rangle$ and $\langle p\rangle$ are the mean values of the position and momentum, respectively (see also Eq.~(\ref{24})). Let us note that this Wigner function is positive definite, as it is also for the harmonic oscillator ground state. This is the reason why the SCS are considered semiclassical states: they remind a classical particle moving cyclically in the oscillator potential with the oscillator period $T=2\pi$ (see Figure (\ref{fig:W10})).

On the other hand, the Wigner functions associated to the {\it even} and {\it odd} coherent states are also obtained from 
their corresponding wavefunctions. We get that
\begin{subequations}
	\begin{eqnarray}
\nonumber W^0_z(q,p)&=&\frac{N_0^2}{\pi} \Big[e^{-(q-\langle q\rangle)^2}e^{-(p-\langle p\rangle)^2}+e^{-(q+\langle q\rangle)^2}e^{-(p+\langle p\rangle)^2}\\
&&+2{\rm Re}\left(e^{-(q+i\langle p\rangle)^2}e^{-(p-i\langle q\rangle)^2}\right)\Big],\label{81} \\
\nonumber W^1_z(q,p)&=&\frac{N_1^2}{\pi} \Big[e^{-(q-\langle q\rangle)^2}e^{-(p-\langle p\rangle)^2}+e^{-(q+\langle q\rangle)^2}e^{-(p+\langle p\rangle)^2}\\
&&-2{\rm Re}\left(e^{-(q+i\langle p\rangle)^2}e^{-(p-i\langle q\rangle)^2}\right)\Big], \label{82}
\end{eqnarray}
\end{subequations}
where the first two terms in both equations correspond to the Wigner functions of the two standard coherent states centered at 
$\pm (\langle q\rangle,\langle p\rangle)$, while the last one is an interference term.

\begin{figure}
	\centering
\includegraphics[scale=0.7]{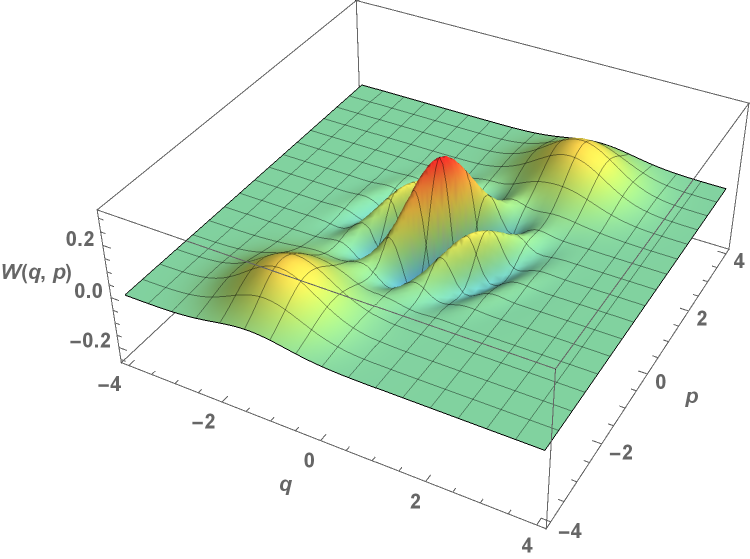}
\caption{Wigner distribution function $W(q,p)$ for an {\it even} coherent state, which shows two separated Gaussian distributions, 
placed in opposite positions in phase space. The oscillations in the middle appear from the interference between the two SCS 
$\vert z\rangle$ and $\vert e^{i\pi}z\rangle$.}\label{fig:W20}
\end{figure}

\begin{figure}
	\centering
\includegraphics[scale=0.7]{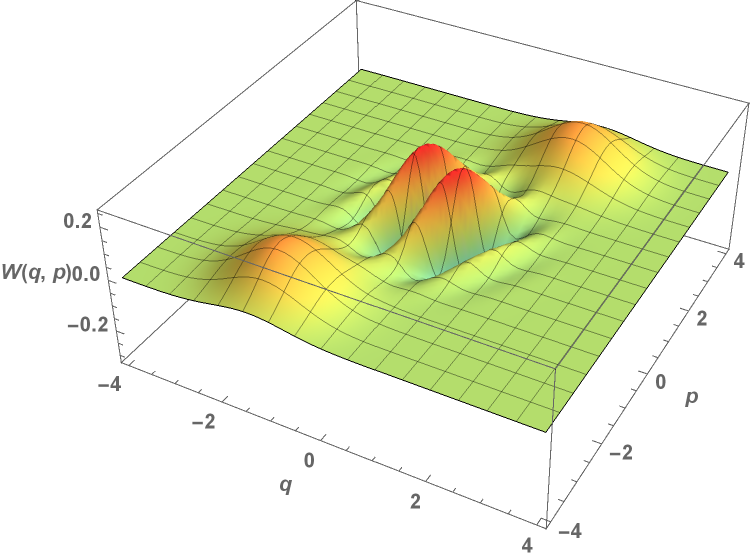}
\caption{Wigner distribution function $W(q,p)$ of an {\it odd} coherent state, which shows two separated Gaussian distributions 
centered in opposite points in phase space. The oscillations in the middle appear from the interference between the two SCS 
$\vert z\rangle$ and $\vert e^{i\pi}z\rangle$.}\label{fig:W21}
\end{figure}

As we can see in Figures \ref{fig:W20} and \ref{fig:W21}, now both Wigner functions take negative values in the region 
between the two Gaussian functions, where the interference term plays a fundamental role. This fact induces to consider them 
as intrinsically quantum states, without any classical counterpart.

Finally, the Wigner functions for the wavefunctions of Eqs.~(\ref{69}-\ref{71}), associated to the triphoton coherent states, turn out to be:
\begin{subequations}
	\begin{eqnarray}
\nonumber W^0_z(q,p)&=&\frac{N_0^2}{\pi} \Bigg[e^{-(q-\langle q\rangle)^2}e^{-(p-\langle p\rangle)^2}+e^{-(q-\langle q\rangle_1)^2}e^{-(p-\langle p\rangle_1)^2}+e^{-(q-\langle q\rangle_2)^2}e^{-(p-\langle p\rangle_2)^2}\\
\nonumber &&+2{\rm Re}\Bigg(e^{-\left(q-\left[\frac{ze^{i2\pi/3}+z^\ast}{\sqrt{2}}\right]\right)^2}e^{-\left(p+i\left[\frac{ze^{i2\pi/3}-z^\ast}{\sqrt{2}}\right]\right)^2}e^{\vert z\vert^2(e^{i2\pi/3}-1)}\\
&&+e^{-\left(q-\left[\frac{ze^{i4\pi/3}+z^\ast}{\sqrt{2}}\right]\right)^2}e^{-\left(p+i\left[\frac{ze^{i4\pi/3}-z^\ast}{\sqrt{2}}\right]\right)^2}e^{\vert z\vert^2(e^{i4\pi/3}-1)} \label{83}\\
\nonumber &&+e^{-\left(q-\left[\frac{ze^{i2\pi/3}+z^\ast e^{-i4\pi/3}}{\sqrt{2}}\right]\right)^2}e^{-\left(p+i\left[\frac{ze^{i2\pi/3}-z^\ast e^{-i4\pi/3}}{\sqrt{2}}\right]\right)^2}e^{\vert z\vert^2(e^{-i2\pi/3}-1)}\Bigg)\Bigg], \\
\nonumber W^1_z(q,p)&=&\frac{N_1^2}{\pi} \Bigg[e^{-(q-\langle q\rangle)^2}e^{-(p-\langle p\rangle)^2}+e^{-(q-\langle q\rangle_1)^2}e^{-(p-\langle p\rangle_1)^2}+e^{-(q-\langle q\rangle_2)^2}e^{-(p-\langle p\rangle_2)^2}\\
\nonumber &&+2{\rm Re}\Bigg(e^{i4\pi/3}e^{-\left(q-\left[\frac{ze^{i2\pi/3}+z^\ast}{\sqrt{2}}\right]\right)^2}e^{-\left(p+i\left[\frac{ze^{i2\pi/3}-z^\ast}{\sqrt{2}}\right]\right)^2}e^{\vert z\vert^2(e^{i2\pi/3}-1)}\\
&&+e^{i2\pi/3}e^{-\left(q-\left[\frac{ze^{i4\pi/3}+z^\ast}{\sqrt{2}}\right]\right)^2}e^{-\left(p+i\left[\frac{ze^{i4\pi/3}-z^\ast}{\sqrt{2}}\right]\right)^2}e^{\vert z\vert^2(e^{i4\pi/3}-1)} \label{84}\\
\nonumber &&+e^{-i2\pi/3}e^{-\left(q-\left[\frac{ze^{i2\pi/3}+z^\ast e^{-i4\pi/3}}{\sqrt{2}}\right]\right)^2}e^{-\left(p+i\left[\frac{ze^{i2\pi/3}-z^\ast e^{-i4\pi/3}}{\sqrt{2}}\right]\right)^2}e^{\vert z\vert^2(e^{-i2\pi/3}-1)}\Bigg)\Bigg], \\
\nonumber W^2_z(q,p)&=&\frac{N_2^2}{\pi} \Bigg[e^{-(q-\langle q\rangle)^2}e^{-(p-\langle p\rangle)^2}+e^{-(q-\langle q\rangle_1)^2}e^{-(p-\langle p\rangle_1)^2}+e^{-(q-\langle q\rangle_2)^2}e^{-(p-\langle p\rangle_2)^2}\\
\nonumber &&+2{\rm Re}\Bigg(e^{i2\pi/3}e^{-\left(q-\left[\frac{ze^{i2\pi/3}+z^\ast}{\sqrt{2}}\right]\right)^2}e^{-\left(p+i\left[\frac{ze^{i2\pi/3}-z^\ast}{\sqrt{2}}\right]\right)^2}e^{\vert z\vert^2(e^{i2\pi/3}-1)}\\
&&+e^{i4\pi/3}e^{-\left(q-\left[\frac{ze^{i4\pi/3}+z^\ast}{\sqrt{2}}\right]\right)^2}e^{-\left(p+i\left[\frac{ze^{i4\pi/3}-z^\ast}{\sqrt{2}}\right]\right)^2}e^{\vert z\vert^2(e^{i4\pi/3}-1)} \label{85}\\
\nonumber &&+e^{i2\pi/3}e^{-\left(q-\left[\frac{ze^{i2\pi/3}+z^\ast e^{-i4\pi/3}}{\sqrt{2}}\right]\right)^2}e^{-\left(p+i\left[\frac{ze^{i2\pi/3}-z^\ast e^{-i4\pi/3}}{\sqrt{2}}\right]\right)^2}e^{\vert z\vert^2(e^{-i2\pi/3}-1)}\Bigg)\Bigg].
\end{eqnarray}
\end{subequations}

\begin{figure}
	\centering
\includegraphics[scale=0.7]{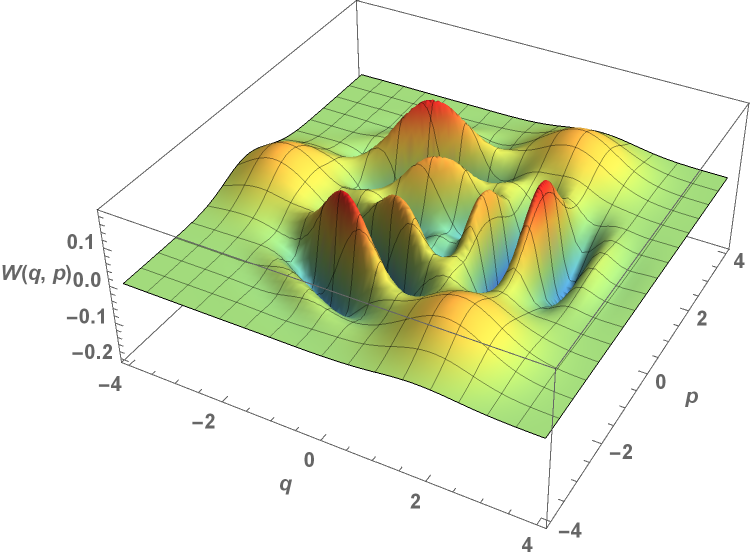}
\caption{Wigner distribution function $W(q,p)$ for a triphoton coherent state $\vert z\rangle_0$, which shows three separated Gaussian distributions, 
placed in the vertices of an equilateral triangle in phase space. The oscillations outside the three vertices appear from the interference between the three SCS $\vert z\rangle$, $\vert e^{i2\pi/3}z\rangle$ and $\vert e^{i4\pi/3}z\rangle$.}\label{fig:W30}
\end{figure}

\begin{figure}
	\centering
	\includegraphics[scale=0.7]{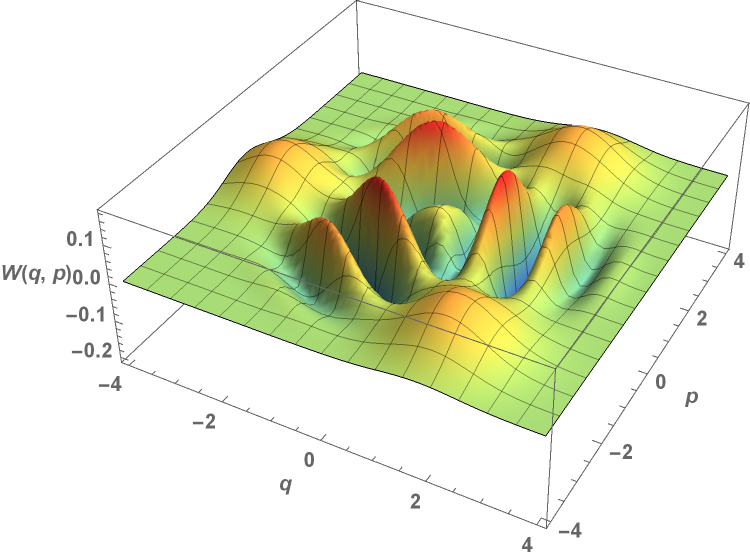}
	\caption{Wigner distribution function $W(q,p)$ for a triphoton coherent state $\vert z\rangle_1$, which shows three separated Gaussian distributions placed 
		in the vertices of an equilateral triangle in phase space. The oscillations outside the three vertices indicate the interference between the three SCS $\vert z\rangle$, $\vert e^{i2\pi/3}z\rangle$ and $\vert e^{i4\pi/3}z\rangle$.}\label{fig:W31}
\end{figure}

\begin{figure}
	\centering
\includegraphics[scale=0.7]{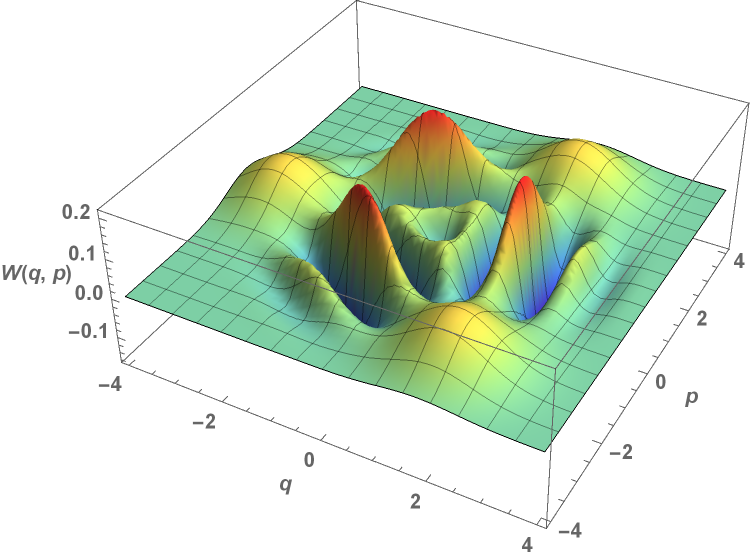}
\caption{Wigner distribution function $W(q,p)$ for a triphoton coherent state $\vert z\rangle_2$, which shows three separated Gaussian distributions placed in the vertices of an equilateral triangle in phase space. The oscillations outside the three vertices reflect the interference between the three SCS $\vert z\rangle$, $\vert e^{i2\pi/3}z\rangle$ and $\vert e^{i4\pi/3}z\rangle$.}\label{fig:W32}
\end{figure}

According to these expressions, each Wigner function $W^j_z(q,p)$, $j=0,1,2$ is composed of the Wigner functions of three standard coherent states centered at the points $(\langle q\rangle,\langle p\rangle)$, $(\langle q\rangle_1,\langle p\rangle_1)$ and 
$(\langle q\rangle_2,\langle p\rangle_2)$ in phase space, together with an interference term. As can be seen in Figures \ref{fig:W30}, 
\ref{fig:W31} and \ref{fig:W32}, the Wigner functions for the triphoton coherent states take negative values, which exhibits clearly the intrinsically quantum nature of these states.

Finally, by substituting Eq.~(\ref{76.1}) in (\ref{77}), the Wigner function $W_z^j(q,p)$ for arbitrary $k$ turns out to be
	\begin{equation}
	W_z^j(q,p)=N_j^2\left[\sum_{s=0}^{k-1}W_{z\mu^{s}}(q,p)+\sum_{s'\neq s=0}^{k-1}\mu^{j(s'-s)}W_{z\mu^{s'},\,z\mu^{s}}(q,p)\right],
	\end{equation}
	where
	\begin{equation}
	W_{\alpha,\,\beta}(q,p)=\frac{1}{\pi}\int_{-\infty}^{\infty}\psi_\alpha^{j\,\ast}(q+y)\psi_\beta^{j}(q-y)\exp\left(2ipy\right)dy,
	\end{equation}
	and $W_{z\mu^{s}}(q,p)$ is the Wigner function associated to the SCS with eigenvalue $z\mu^{s}$ (Eq.~(\ref{80})).

\subsection{Geometric phase}

It is well known that any state $\vert\psi(t)\rangle$ evolving cyclically in the time interval $(0,\tau)$, such that $\vert\psi(\tau) 
\rangle = e^{i\varphi}\vert\psi(0)\rangle$, has associated a {\it geometric phase} $\beta$ \cite{ws89,dpv10,fnos92,fe12}, which depends only of the geometry of the state space (the projective Hilbert space). In particular, if the system is ruled by a time-independent Hamiltonian $H$, the geometric phase is given 
by \cite{fe12,fe94}:
\begin{equation}
\beta =  \varphi + \tau \langle \psi(0) \vert H \vert \psi(0)\rangle.
\end{equation}

\begin{figure}
	\centering
	\includegraphics[scale=0.7]{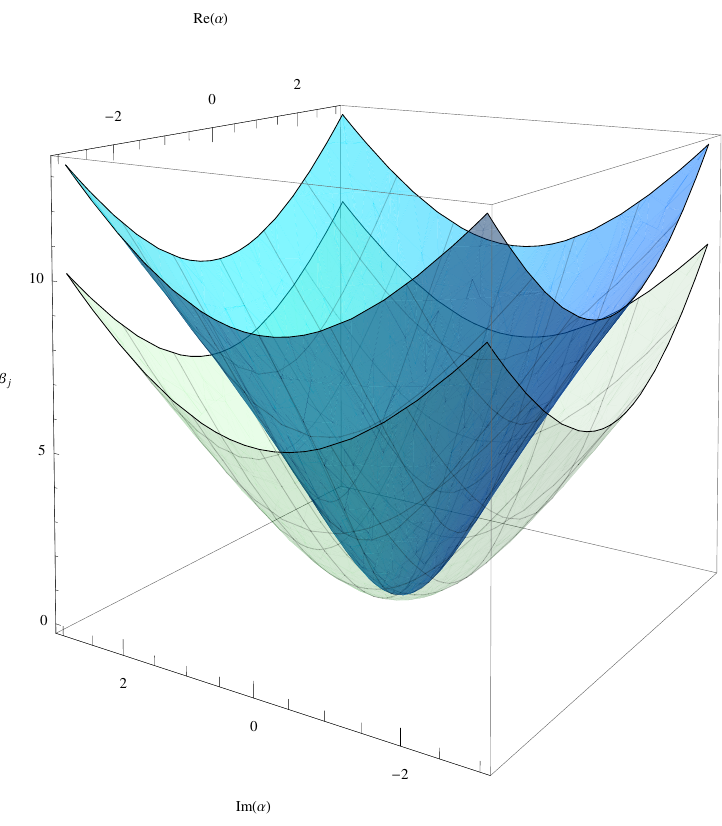}
	\caption{Geometric phase $\beta_j$ as function of $\alpha$ for the {\it even} and {\it odd} coherent states. The cyan graph corresponds to the state $\vert\alpha \rangle_0$, while the gray one to $\vert\alpha\rangle_1$.}\label{fig:geophase2}
\end{figure}
\begin{figure}
	\centering
	\includegraphics[scale=0.7]{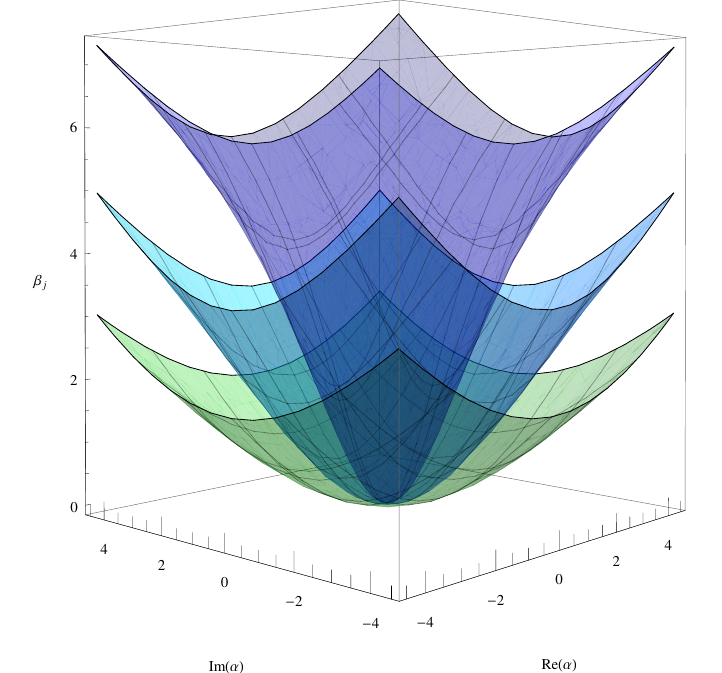}
	\caption{Geometric phase $\beta_j$ as function of $\alpha$ for the triphoton coherent states. The purple graph corresponds to the state $\vert\alpha \rangle_0$, the cyan one to $\vert \alpha\rangle_1$, and the green one to $\vert\alpha\rangle_2$.}\label{fig:geophase3}
\end{figure}

As we saw in the previous sections, the multiphoton coherent states of Eq.~(\ref{22}) evolve cyclically. In fact, by taking $t = \tau = 2\pi/k$
in Eq.~(\ref{37}) it is obtained
\begin{equation}
U(2\pi/k)\vert \alpha \rangle_j = \exp\left(-i\frac{(2j+1)\pi}{k}\right)\vert \alpha \rangle_j.
\end{equation}
Thus, the corresponding geometric phase turns out to be:
\begin{equation}
\beta_j =  -\frac{(2j+1)\pi}{k} + \frac{2\pi}{k} {}_j\langle \alpha \vert H \vert \alpha\rangle_j.
\end{equation}
If we employ the result of Eq.~({33}) we arrive at:
\begin{equation}
\beta_j =  \frac{2\pi}{k}\left(\vert a\vert\alpha\rangle_j\vert^2 - j\right) .
\end{equation}

By taking now $k=2$, {\it i.e.}, for the {\it even} and {\it odd} coherent states, we obtain:
\begin{subequations}
	\begin{eqnarray}
\beta_0&=&\pi \vert \alpha\vert\tanh(\vert\alpha\vert),\\
\beta_1&=&\pi (\vert \alpha\vert\coth(\vert\alpha\vert)-1).
\end{eqnarray}
\end{subequations}
These geometric phases, as functions of $\alpha$, are shown in Figure \ref{fig:geophase2}. We can see that both functions start from the minimum value, $\beta_j=0$, but the geometric phase for the {\it even} states grows more quickly than the one for the {\it odd} states as $\vert\alpha\vert$ grows.

On the other hand, for the triphoton coherent states we arrive at:
\begin{subequations}
	\begin{align}
&\beta_0=\frac{2\pi}{3} \vert\alpha\vert^{2/3}\left(\frac{e^{\vert\alpha\vert^{2/3}}-2 e^{-\frac{\vert\alpha\vert^{2/3}}{2}} \sin \left(\frac{\pi}{6}+\frac{\sqrt{3} \vert\alpha\vert^{2/3}}{2} \right)}{e^{\vert\alpha\vert^{2/3}}+2 e^{-\frac{\vert\alpha\vert^{2/3}}{2}} \cos \left(\frac{\sqrt{3} \vert\alpha\vert^{2/3}}{2}\right)}\right), \\
&\beta_1=\frac{2\pi}{3} \left[\vert\alpha\vert^{2/3}\left(\frac{e^{\vert\alpha\vert^{2/3}}+2 e^{-\frac{\vert\alpha\vert^{2/3}}{2}} \cos \left(\frac{\sqrt{3} \vert\alpha\vert^{2/3}}{2}\right)}{e^{\vert\alpha\vert^{2/3}}-2 e^{-\frac{\vert\alpha\vert^{2/3}}{2}} \sin \left(\frac{\pi}{6} -\frac{\sqrt{3} \vert\alpha\vert^{2/3}}{2}\right)}\right)-1\right], \\
&\beta_2=\frac{2\pi}{3} \left[\vert\alpha\vert^{2/3}\left(\frac{e^{\vert\alpha\vert^{2/3}}-2 e^{-\frac{\vert\alpha\vert^{2/3}}{2}} \sin \left(\frac{\pi}{6}-\frac{\sqrt{3}\vert\alpha\vert^{2/3}}{2}\right)}{e^{\vert\alpha\vert^{2/3}}-2 e^{-\frac{\vert\alpha\vert^{2/3}}{2}} \sin \left(\frac{\pi}{6}+\frac{\sqrt{3}\vert\alpha\vert^{2/3}}{2}\right)}\right)-2\right].
\end{align}
\end{subequations}
These geometric phases, as functions of $\alpha$, are shown in Figure \ref{fig:geophase3}. Once again, we can see that the three functions start from the minimum value, $\beta_j=0$, but the geometric phase for $\vert\alpha\rangle_0$ grows more quickly than those for $\vert\alpha\rangle_1$ and $\vert\alpha\rangle_2$ as $\vert\alpha\vert$ increases.

\section{Conclusions}

In this paper we have discussed the simplest, but very important, realization of the polynomial Heisenberg algebras, generated through 
the harmonic oscillator. In this approach, the generators are the oscillator Hamiltonian $H$ and the ladder operators 
$a^\pm_g=(a^\pm)^k$. By analyzing carefully the algebraic structure that the operator set $\{H, a^-_g, a^+_g\}$ generates, one can obtain the spectrum of $H$ in an unconventional way: it becomes decomposed in a set of $k$ infinity energy ladders, each one of which starts from
one harmonic oscillator eigenstate $\vert\psi^j_{k,0}\rangle$ in the subspace ${\cal H}_{j+1}$. These so-called extremal states, whose number coincides precisely with the order to the differential operator $a^-_g$, become all physical states and from them we can generate all the other 
eigenstates $\vert\psi^j_{k,n}\rangle$ by applying repeatedly $a^+_g$.

We have built as well the corresponding coherent states, as eigenstates of the annihilation operator $a^-_g$ with complex eigenvalue 
$\alpha$. These states are called multiphoton coherent states in the literature, and they have been expressed in terms of the harmonic 
oscillator eigestates $\vert\psi^j_{k,n}\rangle$ belonging to each subspace ${\cal H}_{j+1}$. In general, they are not minimum 
uncertainty states (see Eq.~(\ref{31}) and Figures~\ref{fig:unceraintyk2} and \ref{fig:unceraintyk3}). However, they satisfy a 
{\it partial} completeness relation in each subspace ${\cal H}_{j+1}$ (see Eq.~(\ref{34})). The MCS turn out to be periodic, with a period 
$\tau=2\pi/k$ which is a fraction of the original oscillator one ($T=2\pi$).

The multiphoton coherent states were also expressed in terms of standard coherent states $\vert z\rangle$ with different eigenvalues $z$, as can be seen for some particular values of the integer $k$. For $k=2$ we expanded the {\it even} and {\it odd} coherent states in terms of two SCS with opposite eigenvalues (see Eqs.~(\ref{57}, \ref{58})). On the other hand, for $k=3$ we expressed the triphoton coherent states in terms of three SCS whose labels form an equilateral triangle in the complex plane (see Eqs.~(\ref{66}-\ref{68})). In general, for a MCS $\vert z\rangle_j$ the $k$ standard coherent states involved in 
the superposition are placed uniformly on a circle of radius $\vert z\vert$, each of them having a phase difference of $\phi=2\pi/k$ with the eigenvalues of its neighbor ones, forming the vertices of a regular polygon \cite{d02}. This procedure allowed us to find in a simple way the expressions for the corresponding wavefunctions $\psi^j_z(x)$, as well as their associated Wigner distribution functions $W^j_z(q,p)$. Thus, Figures~\ref{fig:psik2j0}, \ref{fig:psik2j1} and \ref{fig:psik3j0}-\ref{fig:psik3j2} indicate that the probability 
densities evolve cyclically, as Eq.~(\ref{37}) suggests, while Figures~\ref{fig:W20}-\ref{fig:W32} show clearly that the corresponding Wigner functions take negative values. This exhibits the intrinsically quantum nature of the MCS for $k\geq2$, which implies that we cannot use them for semiclassical models, excepting the standard coherent states of Eq.~(\ref{23}) whose semiclassical behavior is very well known. Moreover, due to the cyclic evolutions performed by the multiphoton coherent states, it was natural to calculate their associated geometric phase.

Let us note that the algebraic method addressed in this paper, and several properties of the multiphoton coherent states here generated, appear from the equidistant nature of the harmonic oscillator spectrum, a fact which is also observed in the so-called SUSY harmonic oscillator. Due to this property, it seems natural to try to generate the multiphoton coherent states for the SUSY harmonic oscillator. On the other hand, it would be important to find out if this treatment can be also applied to other systems, whose energy levels are not equally spaced, as well as to their SUSY partner Hamiltonians. Both ideas represent interesting research subjets which are currently under study \cite{df18,fgv18}.

\subsubsection*{Acknowledgments}
The authors acknowledge the support of Conacyt. MCC also acknowledges the Conacyt
fellowship 301117. EDB also acknowledges the hospitality and support of the University of Valladolid.


\begin{thebibliography}{widestlabel}

\bibitem{m84} B Mielnik, J Math Phys \textbf{25} (1984) 3387

\bibitem{vs93} AP Veselov, AB Shabat, Funct Anal App \textbf{27} (1993) 81

\bibitem{dek94} SY Dubov, VM Eleonskii, NE Kulagin, Chaos \textbf{4} (1994) 47

\bibitem{ad94} VE Adler, Physica D \textbf{73} (1994) 335

\bibitem{fhn94} DJ Fern\'andez, V Hussin, LM Nieto, J Phys A \textbf{27} (1994) 3547

\bibitem{as97} N Aisawa, HT Sato, Prog Theor Phys \textbf{98} (1997) 707

\bibitem{fh99} DJ Fern\'andez, V Hussin, J Phys A \textbf{32} (1999) 3603 

\bibitem{acin00} A Andrianov, F Cannata, M Ioffe, D Nishnianidze, Phys Lett A \textbf{266} (2000) 341

\bibitem{cfnn04} JM Carballo, DJ Fern\'andez, J Negro, LM Nieto, J Phys A \textbf{37} (2004) 10349

\bibitem{bf14} D Berm\'udez, DJ Fern\'andez, AIP Conf Proc \textbf{1575} (2014) 50

\bibitem{bfn16} D Berm\'udez, DJ Fern\'andez, J Negro, J Phys A: Math Theor \textbf{49} (2016) 335203

\bibitem{sp95} V Spiridonov, Phys Rev A \textbf{52} (1995) 1909

\bibitem{d02} VV Dodonov, J Opt B \textbf{4} (2002) R1

\bibitem{clm95} O Casta\~{n}os, R Lopez-Pe\~{n}a, VI Man'ko, J Russ Laser Research \textbf{16} (1995) 477

\bibitem{cl12} O Casta\~{n}os, JA L\'opez-Sald\'ivar, J Phys: Conf Ser \textbf{380} (2012) 012017

\bibitem{pk18} OK Pashaev, A Ko\c cak, Springer Proc Math Stat {\bf 266} (2018) 179

\bibitem{kp18} A Ko\c cak, OK Pashaev, arXiv:1812.01842 [quant-ph]

\bibitem{un82} A Ungar, Am Math Mon {\bf 89} (1982) 688

\bibitem{bg71} AO Barut, L Girardello, Comm Math Phys \textbf{21} (1971) 41

\bibitem{bjq90} V Bu\v{z}ek, I Jex, T Quang, J Mod Optics \textbf{37} (1990) 159

\bibitem{b90} V Bu\v{z}ek, J Mod Optics \textbf{37} (1990) 303

\bibitem{mmkw96} C Moore, DM Meekhof, BE King, DJ Wineland, Science \textbf{272} (1996) 1131

\bibitem{bhdmmwrh96} M Brune, E Hagley, J Dreyer, X Ma\^{\i}tre, A Maali, C Wunderlich, JM Raimond, S Haroche, Phys Rev Lett \textbf{77} (1996) 4887

\bibitem{dmm74} VV Dodonov, IA Malkin, VI Man'ko, Physica \textbf{72} (1974) 597

\bibitem{xg89} Y Xia, G Guo, Phys Lett A \textbf{136} (1989) 281

\bibitem{dmn95} VV Dodonov, VI Man'ko, DE Nikonov, Phys Rev A \textbf{51} (1995) 3328

\bibitem{dkmm96} VV Dodonov, YA Korennoy, VI Man'ko, YA Moukhin, Quantum Semiclass Opt \textbf{8} (1996) 413

\bibitem{rr99} B Roy, P Roy, Phys Lett A \textbf{257} (1999) 264

\bibitem{amaj16} D Afshar, A Motamedinasab, A Anbaraki, M Jafarpour, Int J Mod Phys B \textbf{30} (2016) 1650026

\bibitem{cf16} M Castillo-Celeita, DJ Fern\'andez, J Phys: Conf Ser \textbf{698} (2016) 012007

\bibitem{sww91} J Sun, J Wang, C Wang, Phys Rev A \textbf{44} (1991) 3369

\bibitem{cf17} M Castillo-Celeita, DJ Fern\'andez, in {\it Physical and Mathematical Aspects of Symmetries}, S Duarte et al (Eds), Springer, Cham, Switzerland (2017) 111, DOI 10.1007/978-3-319-69164-0

\bibitem{di06} F Dell'Anno, S De Siena, F Illuminati, Phys Rep \textbf{428} (2006) 53

\bibitem{mtk90} A Miranowicz, R Tana and S Kielich, Quantum Opt \textbf{2} (1990) 253

\bibitem{r13} N Riahi, Eur J Phys \textbf{34} (2013) 461

\bibitem{w32} E Wigner, Phys Rev \textbf{40} (1932) 794

\bibitem{kz04} A Kenfack, K Zyczkowski, J Opt B: Quantum Semiclass Opt \textbf{6} (2004) 396

\bibitem{ws89} A Shapere, F Wilczek, {\it Geometric Phases in Physics}, World Scientific, Singapore (1989)

\bibitem{fnos92} DJ Fern\'andez, LM Nieto, MA del Olmo, M Santander, J Phys A \textbf{25} (1992) 5151

\bibitem{dpv10} M Dennis et al, {\it Quantum Phases: 50 years of the Aharonov-Bohm effect and 25 years of the Berry phase}, J Phys A Math Theor \textbf{43} No. 35 (2010)

\bibitem{fe12} DJ Fern\'andez, SIGMA \textbf{8} (2012) 041, DOI: 10.3842/SIGMA.2012.041

\bibitem{fe94} DJ Fern\'andez, Int J Theor Phys \textbf{33} (1994) 2037

\bibitem{df18} E D\'iaz-Bautista, DJ Fern\'andez, Eur Phys J Plus (2019) to be published

\bibitem{fgv18} DJ Fern\'andez, JD Garc\'ia, F Vergara, {\it Multiphoton algebras, coherent states and SUSY partners for a clase of one-dimensional Hamiltonians}, preprint Cinvestav (2018)
\end{thebibliography}
\end{document}